\documentclass[doublecol]{epl2} 

\usepackage[all,2cell]{xy}
\usepackage{color}
\usepackage[latin1]{inputenc}
\usepackage{epsfig}
\usepackage{amsthm}
\usepackage{MnSymbol}
\usepackage{mathrsfs}
\usepackage{nicefrac}
\usepackage{latexsym}
\usepackage{amsfonts}

\newcommand*{\Scale}[2][4]{\scalebox{#1}{$#2$}}%
\newcommand{\tc}[1]{{\color{black} #1}}

\newcommand{\ba}{\begin{array}}
\newcommand{\ea}{\end{array}}
\newcommand{\bes}{\begin{subequations}\bea}
\newcommand{\ees}{\eea\end{subequations}}
\newcommand{\energy}{E}
\newcommand{\W}{W}

\newcommand{\oneto}{\begin{array}{c} 
\xymatrix{ &  \ar@{--}[r]\ar@{--}@<0.15mm>[r]   &  \\ \ar@{->}[ur]\ar@{->}@<0.15mm>[ur] 
 \ar@{<-}[r]\ar@{<-}@<0.15mm>[r] &  \ar@{--}[r]\ar@{--}@<0.15mm>[r]  & &
 \ar@{<-}[ul]\ar@{<-}@<0.15mm>[ul] \ar@{->}[l]\ar@{->}@<0.15mm>[l]   }
\end{array}}

\newcommand{\onefrom}{\begin{array}{c}
\xymatrix{ &  \ar@{--}[r]\ar@{--}@<0.15mm>[r]   &  \\ \ar@{<-}[ur]\ar@{<-}@<0.15mm>[ur] 
 \ar@{->}[r]\ar@{->}@<0.15mm>[r] &  \ar@{--}[r]\ar@{--}@<0.15mm>[r]  & &
 \ar@{->}[ul]\ar@{->}@<0.15mm>[ul] \ar@{<-}[l]\ar@{<-}@<0.15mm>[l]  }
\end{array}}

\newcommand{\twofrom}{\begin{array}{c}
\xymatrix{
\ar@{<-}[dr]\ar@{<-}@<0.15mm>[dr] \ar@{->}[r]\ar@{->}@<0.15mm>[r] &  \ar@{--}[r]\ar@{--}@<0.15mm>[r]  & &
 \ar@{->}[dl]\ar@{->}@<0.15mm>[dl] \ar@{<-}[l]\ar@{<-}@<0.15mm>[l] \\ 
&  \ar@{--}[r]\ar@{--}@<0.15mm>[r]   &   }
\end{array}}

\newcommand{\twoto}{\begin{array}{c}
\xymatrix{
\ar@{->}[dr]\ar@{->}@<0.15mm>[dr]   \ar@{<-}[r]\ar@{<-}@<0.15mm>[r] &  \ar@{--}[r]\ar@{--}@<0.15mm>[r]  & &
 \ar@{<-}[dl]\ar@{<-}@<0.15mm>[dl] \ar@{->}[l]\ar@{->}@<0.15mm>[l] \\ 
&  \ar@{--}[r]\ar@{--}@<0.15mm>[r]   &   }
\end{array}}

\newcommand{\onetwoto}{\begin{array}{c}
\xymatrix{
&  \ar@{--}[r]\ar@{--}@<0.15mm>[r]   &  \\ 
  \ar@{->}[ur]\ar@{->}@<0.15mm>[ur]  \ar@{<-}[dr]\ar@{<-}@<0.15mm>[dr]   &  & &
 \ar@{->}[dl]\ar@{->}@<0.15mm>[dl] \ar@{<-}[ul]\ar@{<-}@<0.15mm>[ul]  \\ 
&  \ar@{--}[r]\ar@{--}@<0.15mm>[r]   &  }
\end{array}}

\newcommand{\onetwofrom}{\begin{array}{c}
\xymatrix{
&  \ar@{--}[r]\ar@{--}@<0.15mm>[r]   &  \\ 
  \ar@{<-}[ur]\ar@{<-}@<0.15mm>[ur]  \ar@{->}[dr]\ar@{->}@<0.15mm>[dr]   &  & &
 \ar@{<-}[dl]\ar@{<-}@<0.15mm>[dl] \ar@{->}[ul]\ar@{->}@<0.15mm>[ul]  \\ 
&  \ar@{--}[r]\ar@{--}@<0.15mm>[r]   &  }
\end{array}}

\title{Carnot efficiency at divergent power output}

\author{Matteo Polettini  \thanks{E-mail: \email{matteo.polettini@uni.lu}}, Massimiliano Esposito} 
\institute{Physics and Materials Science Research Unit, University of Luxembourg, Campus
Limpertsberg, \\ 162a avenue de la Fa\"iencerie, L-1511 Luxembourg (G. D. Luxembourg)} 
 
\date{\today}

\abstract{
The widely debated feasibility of thermodynamic machines achieving Carnot efficiency at finite power has been convincingly dismissed. Yet, the common wisdom that efficiency can only be optimal in the limit of infinitely-slow processes overlooks the dual scenario of infinitely-fast processes. We corroborate that efficient engines at divergent power output are not theoretically impossible, framing our claims within the theory of Stochastic Thermodynamics.  We inspect the case of an electronic quantum dot coupled to three particle reservoirs to illustrate the physical rationale.} 


\begin{document}

\maketitle

\section{Introduction} It is common wisdom that the efficiency $\eta = P_1/ P_2$ of a thermodynamic machine can achieve optimal (or ``Carnot'') efficiency only via quasistatic processes that deliver a vanishing fraction of power output $P_1$ per power input  $P_2$ -- thus making the machine useless for all practical purposes. However, this fact does not appear to be an immediate consequence of the laws of thermodynamics. In fact, the feasibility of useful machines operating at optimal efficiency is an active issue of debate, usually framed  {either in the linear regime (LR), a theory of thermodynamics that describes systems close to equilibrium, or using Stochastic Thermodynamics (ST). The latter assumes an underlying Markovian dynamics of the microscopic degrees of freedom, and encompasses LR by including fluctuations and response far from equilibrium}.

\tc{Humphrey et al. \cite{humphrey} have been among the first to investigate the possibility of Carnot efficiency with finite forces.} In the LR, it has first been argued \cite{benenti}, and then debated \cite{yamamoto}, that asymmetric Onsager coefficients might enhance efficiency at finite power.  Further studies on the feasibility of powerful Carnot efficiency \cite{leggio,ponmurugan} pivot on special assumptions, for example \tc{currents growing less-than linearly (or even discontinuously) in the corresponding forces} \cite{indekeu}. \tc{On the other hand, Carnot efficiency might be impossible to achieve even in the reversible limit when there are leakages at the interface between system and environment \cite{alonso}}. More relevant to our own analysis, in the context of ST, Shiraishi et al. \cite{shiraishi} argued that there is a tradeoff between power and efficiency preventing optimal efficiency at finite power; similar tradeoffs are observed in the study of the maximum efficiency attainable at finite power output \cite{ryabov,holubec,armen,whitney,brandner15,jiang,bauer,proesmans2}.

The latter no-go results appear to be the death knell of all efforts towards optimal and useful machines. Is this search doomed then? We argue that the door is still open to an extreme, yet tantalising, possibility: that the efficiency can be optimised in a regime that is dual and opposite to the quasistatic limit, that of infinitely-fast processes that provide divergent power, yet delivering entropy to the environment at a slower rate. We frame our arguments both in the LR and in the ST of steady-state machines.  The case of the simplest nano-battery charger, viz. a quantum dot (QD) weakly coupled to three electron reservoirs, allows us to illustrate how this limit can be attained by the interplay of strong thermodynamic forces (depths of energy wells) and of fast and slow kinetic parameters (heights of activation barriers).

That achieving powerful optimal efficiency would be \tc{a difficult task} is already clear from the  fact that,  being a rational function,  the efficiency has singular behaviour. For example, in the quasistatic limit the efficiency goes to $0/0$, a fuzzy quantity that can take any value according to how the limit is taken. Hence, it is unreasonable to question what happens ``at optimal efficiency'', and the investigation should rather focus on how limits are approached in certain scaling scenarios towards singularities. The Authors analysed in Ref.\,\cite{polettini} a Gaussian model of efficiency fluctuations where in a regime dubbed {\it singular coupling} the efficiency tends to  $\infty/\infty$, another fuzzy value.
Campisi and Fazio \cite{campisi} noticed that, when the working substance of a quantum Otto cycle consists of $N$ coupled components, in the large-$N$ limit close to a critical point a conspiracy of critical exponents might lead to a super-linear scaling of efficiency versus power. They conclude that ``obstacles hindering the realisation of the critical powerful Carnot engines appear to be of technological nature, rather than fundamental''. Another argument by Shiraishi \cite{shiraishi2} against the attainability of optimal efficiency with finite thermodynamic forces holds provided transition rates are not singular. In a model of an information machine, Bauer et al. \cite{bauer2} found that Carnot efficiency can be achieved for finite cycle times, at infinite precision. Seifert \cite{seifert} has shown that, beyond the LR, there cannot be a universal bound for efficiency at maximum power smaller than Carnot, and that for strong driving the efficiency can be optimised. Raz et al. \cite{raz} ideated engines where maximum power and efficiency are attained at fast driving. Lee and Park observed that Carnot efficiency can be reached at divergent currents in a model of a  Feynman ratchet \cite{park}; Johnson \tc{devised} a Carnot cycle near the critical point in the phase space of a charged black hole, showing that it becomes powerful and efficient at large charge and low pressure and temperature \cite{johnson}. Hence, singular behaviour needs to be inspected more closely. 

\section{Setup and linear regime}  To stage our proposition in a general framework, we hereby  {scale all currents in entropic units of Boltzmann's constant $k_B$ per time}, in such a way that the total power delivered to the environment (dissipation rate) ultimately satisfies the Second Law of thermodynamics:
\begin{align}
\sigma: = P_2 - P_1 \geq 0 .
\end{align}
In this setup the Carnot efficiency is scaled to unity. Setting {\it exactly} $\eta= P_1/P_2  \equiv 1$, the total dissipation rate vanishes $\sigma = 0$, yet the power input and output can be finite. Hence, the Second Law alone bears no consequence on finite power at optimal efficiency. \tc{While the relation between the Second Law and limits on efficiency is debated \cite{brandner,karel,campisi}, let us notice here that we have not yet formulated a constitutive theory describing the behaviour of $P_1$ and $P_2$. As our constitutive theories we will consider LR and ST, but one cannot exclude scenarios where other effective theories might come into play, e.g. when the working substance is close to a critical point \cite{indekeu}.}

The First Law enters the scene when power is resolved into conserved units of energy (thermodynamic forces $\vec{F}$), and of velocities of their carriers (thermodynamic currents $\vec{J}$):
\begin{equation}
\begin{aligned}
P_1 & =  - F_1 J_1 ,  & & & 
P_2 & =  F_2 J_2 \geq 0.
\end{aligned} 
\end{equation}
The negative sign for $P_1$ highlights that $J_1$ is expected to perform work {\it against} the corresponding force, which is the ultimate purpose of machines at all scales: to lift weights against gravity, to transduce molecules across membranes against osmotic pressure, etc.

We now need a \tc{constitutive} theory for the relation between currents and forces. The simplest theory is LR, prescribing the constitutive relations $\vec{J} = \mathbb{L} \, \vec{F}$  with positive-semidefinite response matrix $\mathbb{L} = (L_{ij})_{i,j}$, on the assumption that $L_{12}L_{21} \geq 0$. The efficiency can be expressed as \cite{entin}
\begin{align}
\eta_{\mathrm{LR}} 
 & = - \frac{1 +  \vartheta \, \phi \, \sqrt{1 - \zeta}}{ \phi^2 +  \phi/\vartheta \,\sqrt{1 - \zeta}}
\end{align}
in terms of only three adimensional parameters, namely
\begin{align}
\vartheta := \sqrt{\frac{L_{12}}{L_{21}}}, \quad
\zeta := \frac{\det L}{L_{11}L_{22}}, \quad
\phi := \frac{F_2}{F_1} \sqrt{\frac{L_{22}}{ L_{11}}} . 
\end{align}
This compact form is convenient for studying the functional behaviour of the efficiency. The first two parameters are ``structural'', as they characterize the apparatus: the first measures the violation of Onsager's symmetry; the second is related to the so-called figure of merit  \cite{colloqium}. The third parameter is ``contingent'', meaning that it depends explicitly on the applied forces, the ``knobs'' that \tc{an} ideal observer handles  {(the condition $P_2 \geq 0$ imposes $\phi \leq - \vartheta^{-1}\sqrt{1-\zeta} \lor \phi \geq 0$)}. Optimising the efficiency with respect to the latter parameter $\phi$ we obtain
\begin{align}
\eta^\star_{\mathrm{LR}} := \mathrm{sup}_\phi \, \eta_{\mathrm{LR}} = \vartheta^2 \frac{1 - \sqrt{\zeta}}{1 + \sqrt{\zeta}}, \label{eq:sup}
\end{align}
reached at $\phi^\star = -(1+\sqrt{\zeta})/\vartheta\sqrt{1-\zeta})$ within the above-mentioned interval.

For the sake of generality, so far we allowed for asymmetric response coefficients, that apparently can be exploited to manipulate the efficiency at will. However, the quantification of dissipation in models with odd dynamical variables is debated. \tc{In the context of the ST of autonomous machines with two terminals (input/output) symmetry of response coefficients is a structural property. To obtain non-symmetric response coefficients, one needs to employ to a larger multi-terminal model and then require that the extra currents vanish \cite{yamamoto,brandner13}. An analysis of the effect of these ``stalling currents'' \cite{altaner} indicates that Carnot efficiency cannot be achieved. However, asymmetric Onsager coefficients can be achieved in time-periodic machines \cite{karel}. Let us here assume} that time reversal symmetry holds at a fundamental level and set $\vartheta =1$ in the following.

\tc{Optimal efficiency then only depends on parameter $\zeta$}. From Eq.\,(\ref{eq:sup}) it is clear that Carnot efficiency can only be achieved when $\zeta \to 0$ {\it after} $\phi$ has been optimised. \tc{In Ref.\,\cite{polettini} we showed that limits in $\zeta$ and limits in $\phi$ do not commute, and therefore one can approach the fuzzy limit of vanishing power input/output with any value of the efficiency, even the completely ``dud'' machine with $\eta = -1$. This shows how delicate efficiency optimisation is and in what sense $0/0$ is a fuzzy value}. A vanishing $\zeta$ is related to a high figure of merit. It is often assumed that the so-called tight-coupling limit $\zeta \to 0$ only occurs when the determinant of the linear-response matrix vanishes
$\det \mathbb{L} \to 0$. However, $\zeta \to 0$ is more generally achieved when the determinant is \tc{much smaller than} the product of the diagonal response coefficients $L_{11}L_{22}$, a limit that was briefly analysed by the Authors in Ref.\,\cite{polettini}.

\section{Far from equilibrium}  We now  venture far from equilibrium, turning to steady-state ST. As the simplest possible system we consider two states ($l$ and $r$)  {occupied with probability $p_{l/r}$}, among which three distinguishable transitions $i = 0,1,2$ can occur  {at rates $w^{\pm}_i$ rates of jumping right or left ($+/-$)}
\begin{align}
\xymatrix{l \ar@{-}[r]^0 \ar@{-}@/_{15pt}/[r]_2  \ar@{-}@/^{15pt}/[r]^1 & r }. \label{eq:twostate}
\end{align}
 {The current \tc{from one state to the other} $J_i = w_i^+ p_l - w_i^- p_r$ counts the net number of transitions. The two fundamental constraints ruling network thermodynamics \cite{oster} \tc{are} Kirchhoff's Loop and Current Laws. \tc{They can be implemented using cycle analysis} \cite{schnak,polettiniprojectors}, which states that independent contributions to the dissipation rate are described by the set of fundamental cycles}\begin{align}
 \ba{c}\xymatrix{l \ar@{<-}[r]^0 \ar@{->}@/_{15pt}/[r]_2  & r } \ea, \qquad \ba{c}\xymatrix{l \ar@{<-}[r]_0  \ar@{->}@/^{15pt}/[r]^1 & r }\ea,
\end{align}
respectively related to the intake of power from reservoir 2 and the outtake of power by reservoir 1, with respect to reservoir 0.  {Defining $\W_{ij} := w_{i}^+ w_{j}^-$ and  $\W^\pm_{ij} := \W_{ij} \pm \W_{ji}$,} we obtain for  the currents and forces 
\begin{equation} 
\begin{aligned} \label{eq:J}
J_1 & = \frac{Z_{10}\W^-_{10} + Z_{12} \W^-_{12}}{Z}, & F_1 & = \log \frac{\W^+_{10}+\W^-_{10}}{\W^+_{10}-\W^-_{10}}, \\
J_2 & = \frac{Z_{20}\W^-_{20} - Z_{12}\W^-_{12}}{Z} , & F_2  & = \log \frac{\W^+_{20}+\W^-_{20}}{\W^+_{20}-\W^-_{20}},
\end{aligned}
\end{equation}
where $Z:= \sum_i (w^{+}_i + w^{-}_i)$ is a normalisation factor associated to the spanning-tree expression of the steady-state probability of being either left or right \cite{schnak,zia}. The coefficients $Z_{ij}$, to be defined later for the sake of greater generality, are unity for the present system, $Z_{ij} = 1$. Finally the efficiency reads
\begin{align}
\eta & = \frac{(Z_{12} \W^-_{12} + Z_{10}  \W^-_{10}) F_1}{ (Z_{12} \W^-_{12} - Z_{20} \W^-_{20}) F_2}. \label{eq:efficiency}
\end{align}
Notice that (for $Z_{ij} =1$) the efficiency only depends on five dimensional parameters, which can be made into four independent adimensional parameters, two more than symmetric LR, reflecting the fact that far from equilibrium time-symmetric dynamical properties of the system play a crucial role \cite{baiesi}. The LR  can eventually be recovered when $\W^-_{ij} \ll \W^+_{ij}$, yielding the response matrix
\begin{align}
\mathbb{L} = \frac1{2Z}\left( \ba{cc}
Z_{10}\W^+_{10} + Z_{12}\W^+_{12}  &  - Z_{12}\W^+_{12} \\
- Z_{12}\W^+_{12}  & Z_{20}\W^+_{20} + Z_{12}\W^+_{12}
\ea \right), \label{eq:lrm}
\end{align}
satisfying Onsager's symmetry. The LR efficiency has the simple expression
\begin{align}\eta_{\mathrm{LR}} & = \frac{(Z_{12} \W^-_{12} + Z_{10}  \W^-_{10})\W^-_{10}\W^+_{20}}{(Z_{12} \W^-_{12} - Z_{20} \W^-_{20})\W^-_{20}\W^+_{10}}. \end{align}

\tc{But}, let us stay far away from equilibrium. We notice that in the limit where $|\W^-_{12}|$ is much greater than all other $|\W^-_{ij}|$'s, the efficiency goes to
\begin{align}
\eta \to  
&  F_1/ F_2. 
\end{align}
A very large negative $\W^-_{12} \to - \infty$ also implies $w^+_1w_2^- \ll w^-_1w_2^+$, which with a little cycle algebra can be recast as
\begin{align}
F_2 - F_1 \to \infty. \label{eq:F}
\end{align}
\begin{figure}
\centering
\includegraphics[width=0.95\columnwidth]{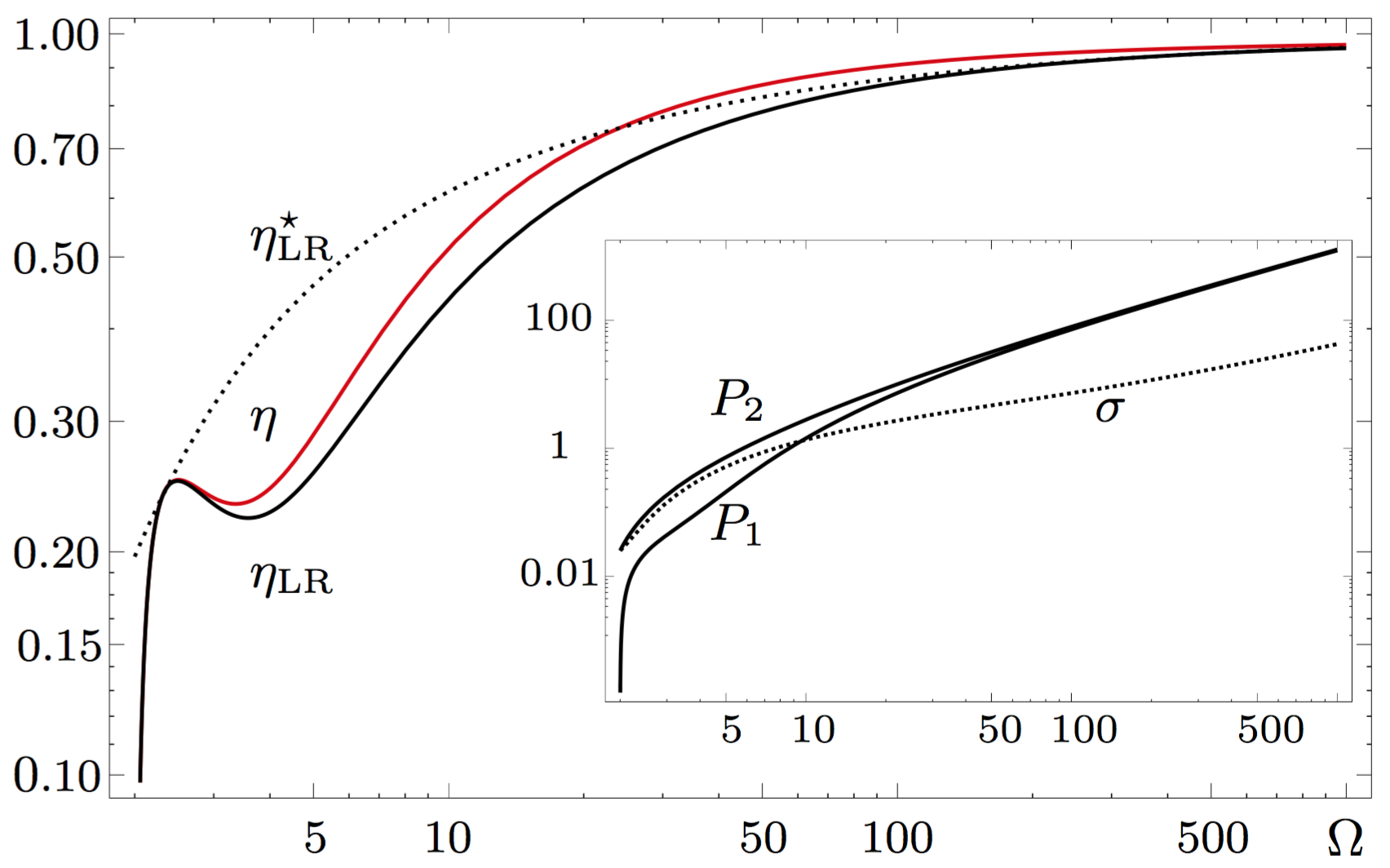} \caption{The efficiency $\eta$ (upper red curve), its linear-regime approximation $\eta_{\mathrm{LR}}$, and the maximal linear-regime efficiency $\eta^\star_{\mathrm{LR}}$ (dotted) as functions of the scaling parameter, in logarithmic scale, for $\alpha = \delta = 1$, $\beta = 2$, and with choice of rates (at $\Omega=1$)
$w _0^{\pm} = w_2^- = w_1^- = 1$, $w_1^+= 0.5$, $w_2^+= 0.6$. Inset: Log-log plot of the power input and output as a function of the scaling parameter. The dotted line is the difference between the two, showing that the dissipation stays order of magnitudes smaller than the power input and output. In fact, by sending $w_1^+/w_2^- \to 1$ (at $\Omega =1$), one can make the slope of the curve of the dissipation lower at will.}
\label{fig:efficiency}
\end{figure}
The last two equations are the focal point of our discussion. They highlight that, in order to achieve optimal efficiency, the difference between the affinities should grow large, but at the same time we don't want their ratio to decay to zero. Thus, \tc{in our framework we observe that} thermodynamic forces cannot be finite for optimal efficiency to be achieved. However, our two focal equations also suggest \tc{that} optimal efficiency can be achieved in a scaling scenario where both forces grow indefinitely, but proportionally. One scaling scenario that realises this limit is given by $w _0^{\pm} \sim 1$, $w_2^- \sim \Omega^{\alpha}$, $w_2^+ = \Omega^{\alpha+\delta}$, $w_1^- \sim \Omega^\beta$, $w_1^+ \sim \Omega^{\beta+\delta}$, with $\beta > \alpha > 1$ and $\delta > 0$.
In such scenario, in Fig.\,\ref{fig:efficiency} we plot the efficiency and the power input/output as functions of $\Omega$, clearly showing that the singular limit exists.  It can be shown that in this limit the LR parameters $\zeta \to 0$ and $\phi \to 1$. Diagrammatically, it can be interpreted as a prevalence of circulation around cycle
$\xymatrix{ \ar@{->}@/_{7pt}/[r]|-2  \ar@{<-}@/^{7pt}/[r]|-1 & }$.
 
\section{Fluctuations} 
\begin{figure}
  \centering
  \includegraphics[width=0.95\columnwidth]{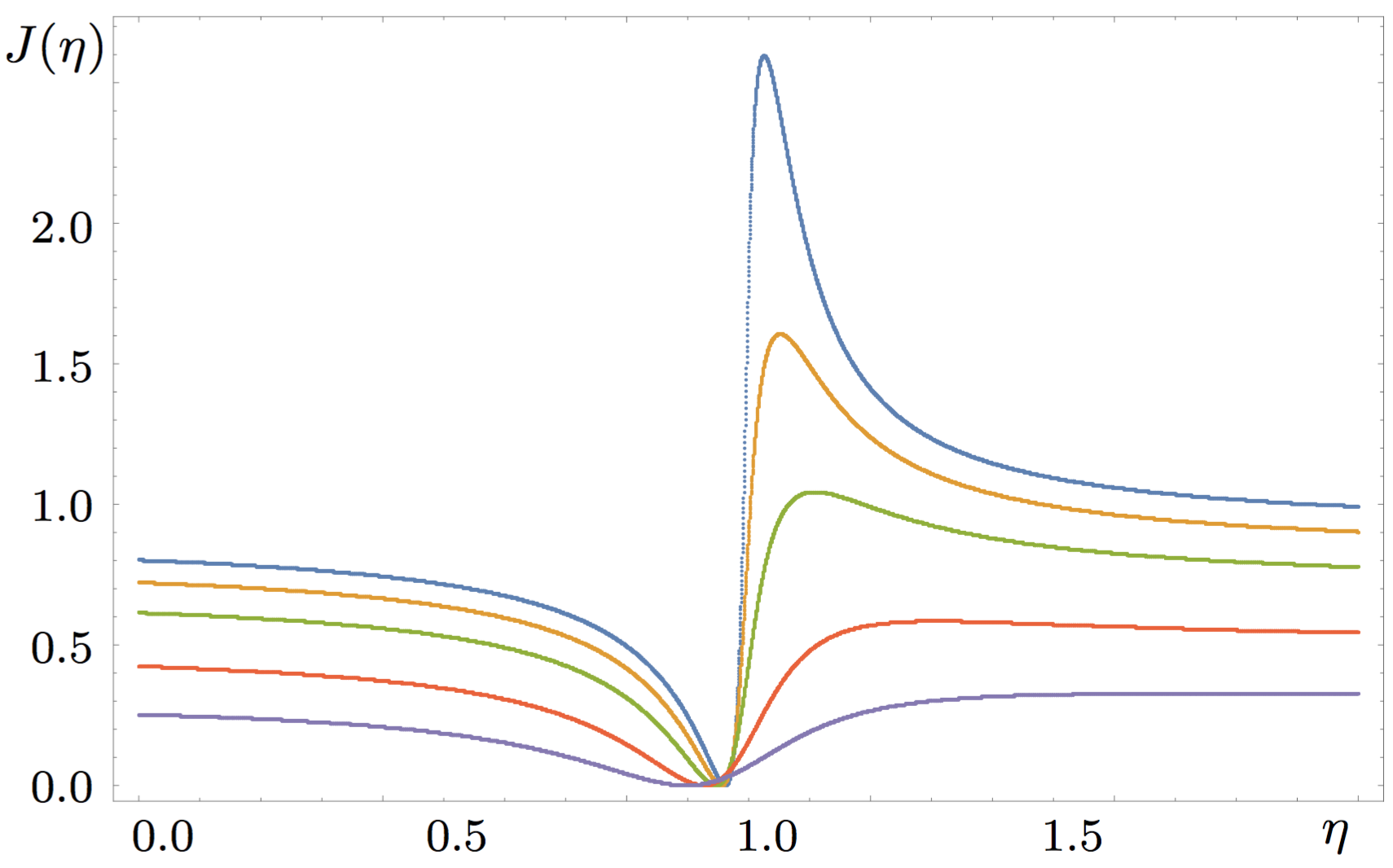} 
\caption{The rate function for the efficiency, for values $\Omega = 10,20,50,100,200$, from lower to upper curve.}
\label{fig:rate}
\end{figure}

So far we have considered the average behaviour of the currents. In the context of ST observables fluctuate. Efficiency as a stochastic variable has first been studied by Verley et al. \cite{verley1}, leading to system-specific analysis  \cite{proesmans1} and to a general characterisation of its universal features \cite{verley2}. At the steady state, the probability $P_t(\eta)$ of achieving efficiency $\eta$ at time $t$ is described by the rate function $J(\eta) = -\lim_{t \to \infty} \frac1{t} \log P_t(\eta)$. In general, rate functions of stochastic variables are Legendre-dual to time-scaled cumulant generating functions, which are often more tractable. However, the efficiency has no finite moments \cite{polettini,proesmans1}, hence its cumulant generating function is nowhere defined. Yet the formula by Verley $J(\eta) = - \min_{q} \lambda(\eta q,q)$ allows to obtain the efficiency's rate function from the the scaled-cumulant generating function $\lambda(q_1,q_2)$ of the currents, up to a constant. The two-state model is analytically tractable to a great extent. In Fig.\,\ref{fig:rate} we plot the rate function of the efficiency for several values of the scaling parameter $\Omega$. The most probable efficiency, given by the infimum of the rate function, slowly creeps to Carnot. The peak, representing the most unlikely value of the efficiency \cite{verley1}, also approaches Carnot. Hence, in the singular limit, Carnot efficiency is at the same time the most and the least probable value. It then follows that in the large-$\Omega$ limit small fluctuations might lead to dramatic effects.

\section{Example: Quandum Dot} 
\begin{figure}
  \centering
  \includegraphics[width=0.95\columnwidth]{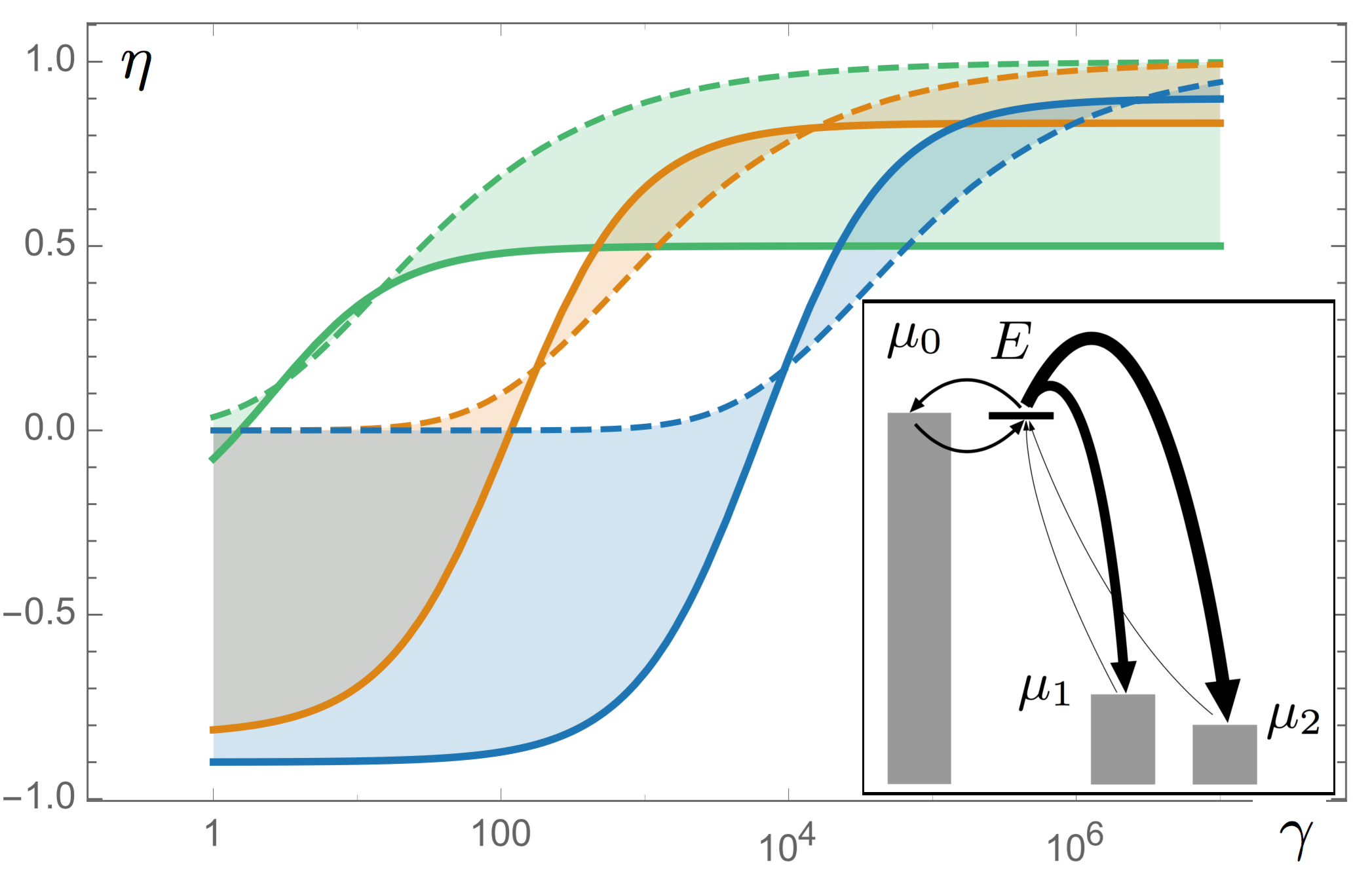} 
\caption{{\bf Inset}: Illustration of the transitions between the QD and the reservoirs at different chemical potentials. {\bf Main frame}: Thick curves, from left to right, are log-linear plots of the efficiency as a function of the \tc{relative} tunnelling rate $\gamma=\gamma_1=\gamma_2$, for $\mu_1 - \mu_2 = k_B T$, and for optimal efficiencies $\eta_{\mathrm{opt}}=.5,.83,.9$ (see Eqs.\,(\ref{eq:effeff},\ref{eq:powpow}). The corresponding dotted lines (separated by shadings) represent the optimal efficiencies $\eta^\star_{\mathrm{LR}}$ that a machine operating in the LR can achieve for the same values of tunnelling rates.}
\label{fig:QD}
\end{figure}

The above two-level system could be physically realised as a single electronic QD acting as an energy filter for electron transport between three reservoirs at the same temperature $T$, held at different chemical potentials $\mu_i$ by the action of external electric fields \cite{mahan,svilans,humphrey,esposito}. Such device can be seen as a work-to-work conversion nanomachine (a nano-battery charger). The QD can either be ($l$) occupied at fixed energy $\energy$  or ($r$) empty at zero energy. Letting $f(\delta):=(1+e^\delta)^{-1}$ be the Fermi distribution, the QD is charged by the $i$-th reservoir  at rate $w_i^{-} = \gamma_i f(\delta_i)=: \gamma_i f_i$  and discharged at rate $w_i^+ = \gamma_i (1-f_i)$, where $\gamma_i$ is the tunnelling rate, and $\delta_i := (\energy-\mu_i) / k_B T$ (see inset of Fig.\,\ref{fig:QD} for an illustration and the Appendix for further details). We set $\gamma_0 \equiv 1$, and $\mu_0 \equiv \energy$, which means that the reference reservoir has no bias, $f_0 = 1- f_0 = 1/2$. In view of Eqs.\,(\ref{eq:J},\ref{eq:efficiency}) the expressions for the efficiency and the power output are given by
\begin{align}
\eta & = \eta_{\mathrm{opt}}  \;\; \frac{\gamma_1 \gamma_2  \left( f_1 - f_2\right) + 
  \gamma_1  \left( f_1 - 1/2\right)}{\gamma_1 \gamma_2  \left( f_1 - f_2\right) - \gamma_2   \left( f_2 - 1/2\right)},  \label{eq:effeff} \\
P_1 & = \delta_1 \; \;\frac{\gamma_1 \gamma_2  \left( f_1 - f_2\right) + 
  \gamma_1 \left( f_1 - 1/2\right)}{1 + \gamma_1 + \gamma_2} , \label{eq:powpow} 
\end{align}
where
$\eta_{\mathrm{opt}} =  \delta_1/ \delta_2$. We choose for reservoirs $1$ and $2$ extremely biased rates in favour of discharging the dot $\delta_2 \gnsim \delta_1 \gg 0$. Large tunnelling rates then allow to asymptotically reach any given value of the optimal efficiency $\eta_{\mathrm{opt}} $ (see Fig.\,\ref{fig:QD}), boosting the power output. Comparison with the optimal efficiency of the LR scenario at small chemical potential differences $\epsilon_i  = (\mu_i - \mu_0)/ k_B T$ indicates that infinitely fast and slow processes might be afflicted by the very same technological issues.

\section{Generalisation and discussion} Let us now show that the above arguments generalise to more complex models in ST consisting of a large state space, but with only two cycles.  {The state-space is depicted by a graph}
\begin{align}
\ba{c} \Scale[0.7]{
\xymatrix{
& \bullet \ar@{--}[r] & \bullet \\
\bullet	 \ar@{-}[ur]	 \ar@{-}[dr] \ar@{-}[r] 
& \bullet \ar@{--}[r] & \bullet &
\bullet  \ar@{-}[ul]	 \ar@{-}[dl] \ar@{-}[l]   \\
& \bullet \ar@{--}[r]  & \bullet 
}}\ea , \quad \mathrm{where\;e.\;g.} \quad   
\bullet = \ba{c}  \Scale[0.5]{ \xymatrix{  
 &  &   \ar@{-}[d]  \\
 \ar@{-}[u]  & \ar@{-}[l] \ar@{-}[u]   \ar@{-}[dr]    \ar@{-}[ur]    &  \\
 \ar@{-}[ur] &   & 
 }} 
 \ea .
\end{align}
 {Each bullet in this diagram denotes a so-called {\it arborescense} of the graph, i.e. subgraphs that do not contain cycles and that do not have sites in common among themselves. Whatever the ramification of the arborescences, indeed the system only has two cycles.} Dashed lines denote an arbitrary number of edges connecting the two bullet sites  {at their extremes} via other bullet sites.  The theory of Hill \cite{hill} for steady-state currents leads to the expression Eq.\,(\ref{eq:efficiency}) where
\begin{equation}
\begin{aligned}
\W_{10} & = \Scale[0.6]{\oneto},  \\
\W_{20} & = \Scale[0.6]{\twoto}, \\
\W_{12} & = \Scale[0.6]{\onetwoto}.
\end{aligned}
\end{equation}
Here, the diagrammatic expression stands for the product of rates along the oriented links of the diagram,  {with $\W_{01}, \W_{02}, \W_{21}$ given by the diagrams with inverse orientation.} The coefficients $Z_{ij}$ can be obtained as the oriented and rooted spanning tree polynomials in the network where the corresponding cycle is contracted to a unique site \cite{polettiniphd,zia}. This given, the above treatment follows unchanged from Eq.\,(\ref{eq:J}) to Eq.\,(\ref{eq:F}).

\tc{ As regards the bound between the efficiency and power output proven by Shiraishi \cite{shiraishi}, we notice that the relation is mediated by a dimensional factor, which is needed to compensate for the fact that while the efficiency is adimensional, power has physical units, hence it must be measured against some dynamical property of the system}.  While a direct application of their result to our setting is not possible, we can retrace their derivation in its fundamental steps \footnote{See the Appendix for details. Notice in particular that Shiraishi et al. considered weighted currents supported on several edges; this requires several additional passages, including a tilting of the rates, and the application of the Schwarz inequality.}. After some work we obtain
\begin{align}
\eta(1-\eta) \geq \frac{8}{9} \frac{P_1}{K_2 F_2^2} \label{eq:shira}
\end{align}
where
\begin{align}
K_2 = \frac1{Z} (\W^+_{20} + \W^+_{12}).
\end{align}
The denominator in Eq.\,(\ref{eq:shira}) can be interpreted as a sort of nonequilibrium LR approximation of the power input, and it is not universal. In particular, in our simple model it is easily seen to be systematically larger \tc{than} the power output by orders of magnitude. Therefore our treatment is consistent with previous claims of trade-offs between efficiency and finite power.
\tc{Recently another bound involving power, efficiency and the power's variance has been found by Pietzonka and Seifert \cite{pietzonka}, which seems to suggest that in an efficient engine with divergent power, the power's variance should also diverge, another fact that is vaguely reminiscent of phase transitions, as already suggested in Refs. \cite{polettini,campisi}.}

Our analysis is restrained to two-cycle steady-state (autonomous) models in ST, leaving out more complicated models of multiterminal machines with stalled currents, of multicyclic systems whereby symmetries imply that to the same physical process there contribute many configuration cycles \cite{polettini2016}, and of time-dependent processes. We mention, without further discussion, that in networks of coupled overdamped oscillators described by Langevin-type equations a similar type of analysis holds \cite{polettiniwip}. All of the above-mentioned setups rely on the Markov assumption, which in the strong driving limit might fail.

\tc{To conclude, our study clearly shows that the quest for optimal efficiency in machines well-described by ST leads in two directions: the reversible and the deeply irreversible limits. While the first is widely known and of limited practical purposes, the second one has been systematically overlooked. We used a simple system to provide a proof-of-concept of the existence of this second limit, and we characterised it physically in terms of the electronic transport across a Quantum Dot. This research might  lead to better design principles for powerful and efficient machines.}

\acknowledgments

Discussion with J. O. Indekeu and, over the years, with G. Verley was very stimulating.
The research was supported by the National Research Fund Luxembourg (Project FNR/A11/02) 
and by the European Research Council (Project No. 681456).

\section{Appendix: Power/efficiency tradeoff}

We provide a derivation of the power/efficiency tradeoff relation in our context. Let $w_{ij}$ be the rates of a Markov jump process, and $p_j$ the steady-state probability. 
We follow the derivation of Shiraishi et al. [Phys. Rev. Lett. {\bf 117}, 190601 (2016)]. First, as observed by Whitney in Ref.\,[Phys. Rev. Lett. {\bf 112}, 130601 (2014)],
\begin{align}
\eta (1 - \eta) & = \frac{P_1}{P_2} \left(1 - \frac{P_1}{P_2} \right) = \frac{P_1 \sigma}{P_2^2} 
\end{align}
Then, using Shiraishi's Eq.\,(16) and along the following treatment, the steady-state dissipation rate can be expressed as
\begin{align}
\sigma = \sum_{i \neq j} s\left(w_{ij}p_j,w_{ji}p_i\right)
\end{align}
where
\begin{align}
s\left(a,b\right) := a \log \frac{a}{b} + a - b.
\end{align}
The following inequality holds
\begin{align}
s\left(a,b\right) \geq \frac{8}{9} \frac{(a - b)^2}{a + b}.
\end{align}
Hence we obtain
\begin{align}
\sigma \geq \frac{8}{9} \sum_{i \neq j}  \frac{J_{ij}^2}{K_{ij}}.
\end{align}
where
\begin{align}
K_{ij} = w_{ij} p_j + w_{ji} p_i.
\end{align}
In our example:
\begin{align}
\sigma & \geq \frac{8}{9} \left(  \frac{J_1^2}{K_1} + \frac{J_2^2}{K_2} + \frac{J_0^2}{K_0} \right) \label{eq:bound} 
\end{align}
and in particular 
\begin{align}
\sigma & \geq \frac{8}{9}  \frac{P_2^2}{K_2 F_2^2} 
\end{align}
yielding
\begin{align}
\eta (1 - \eta) & \geq \frac{8}{9} \frac{P_1}{K_2 F_2^2}.
\end{align}
Notice that if instead of throwing all terms away in Eq.\,(\ref{eq:bound}) we express $J_0 = J_1 - J_2$, we can obtain a slightly more accurate inequality
\begin{align}
\eta (1 - \eta) & \geq  P_1 \left(  \frac{8}{9} \frac{1}{K_{\mathrm{eff}} F_2^2}- \eta \frac{16}{9}\frac{1}{K_{0} F_1 F_2}\right).
\end{align}

\section{Appendix: Quantum dot}

We provide a more complete treatment of the single Fermionic quantum dot coupled to three reservoirs. Each reservoir charges and discharges the quantum dot at rate
\begin{subequations}
\begin{align}
w^{+}_i & = \gamma_i  f(\delta_i) \\
w^{+}_i & = \gamma_i  \left[ 1- f(\delta_i)\right]
\end{align} 
\end{subequations}
where
\begin{align}
f(\delta) = \frac{1}{e^{\delta} + 1}
\end{align}
is the Fermi distribution and $\gamma_i$ is the tunnelling rate from the reservoir to the quantum dot. Furthermore,
\begin{align}
\delta_i = \frac{\energy- \mu_i}{k_B T},
\end{align}
where $T$ is the temperature of all reservoirs, $\epsilon$ the energy difference between the charged and uncharged quantum states, and $\mu_i$ is the chemical potential of the $i$-th reservoir. The Fermi distribution satisfies the Kubo-Martin-Schwinger (local detailed balance) condition
\begin{align}
\frac{f(\delta)}{1-f(\delta)} = e^{-\delta}.
\end{align}
For our relevant quantities in the determination of the efficiency we obtain
\begin{subequations}
\begin{align}
W_{ij}^{-} & = \gamma_i \gamma_j  \left[ f(\delta_i) - f(\delta_j)\right], \\
W_{ij}^{-} & = \gamma_i \gamma_j  \left[ f(\delta_i) + f(\delta_j) - 2 f(\delta_i) f(\delta_j) \right],
\end{align}
\end{subequations}
and for the thermodynamic forces
\begin{align}
F_i & = \delta_0 - \delta_i = \beta(\mu_i - \mu_0),
\end{align}
yielding the efficiency
\begin{align}
\eta & = \frac{\gamma_1 \gamma_2  \left[ f(\delta_1) - f(\delta_2)\right] + 
  \gamma_1 \gamma_0  \left[ f(\delta_1) - f(\delta_0)\right]}{\gamma_1 \gamma_2  \left[ f(\delta_1) - f(\delta_2)\right] - \gamma_2 \gamma_0  \left[ f(\delta_2) - f(\delta_0)\right]} \; \cdot\;\frac{\delta_1-\delta_0}{\delta_2-\delta_0}.
\end{align}
Remarkably, the normalisation factor $Z$ simplifies to 
\begin{align}
Z = \gamma_0 + \gamma_1 + \gamma_2.
\end{align}
The power input and output are then given by
\begin{subequations}
\begin{align}
P_1 & = \gamma_0 (\mu_1 - \mu_0) \frac{\gamma_1 \gamma_2  \left[ f(\delta_1) - f(\delta_2)\right] + 
  \gamma_1 \left[ f(\delta_1) - f(\delta_0)\right]}{1 + \gamma_1 + \gamma_2} \\
P_2 & = \gamma_0 (\mu_2 - \mu_0) \frac{\gamma_1 \gamma_2  \left[ f(\delta_1) - f(\delta_2)\right] - \gamma_2 \left[ f(\delta_2) - f(\delta_0)\right]}{1 + \gamma_1 + \gamma_2}  
\end{align}
\end{subequations}
In the main text, we further simplified the analysis by setting the timescale $\gamma_0 = 1$, by assuming that the tunnelling rates of the other reservoirs are $\gamma_1 = \gamma_2 = \gamma$, by setting $\delta_0 = 0$ (that is, $\mu_0 =\energy$) and by setting $\mu_1-\mu_2 = k_B T$ (that is, $\delta:= \delta_1  = \delta_2 - 1$). We then obtain the function plotted in Fig.\,3:
\begin{align}
\eta(\gamma,\delta) & = \frac{\gamma \left[ f(\delta) - f(\delta+1)\right] + 
  \left[ f(\delta) -1/2\right]}{\gamma   \left[ f(\delta) - f(\delta+1)\right] - \left[ f(\delta+1) - 1/2\right]} \; \cdot\;\frac{\delta}{\delta +1}.
\end{align}
The linear regime expression used to plot the maximum theoretical efficiency in Fig.\,3 is obtained by assuming small thermodynamic forces, i.e. by expanding $\delta_i = \delta_0 + \epsilon_i$. Using $f(\delta_i) = f(\delta_0) + f'(\delta_0)\epsilon_i$ one obtains the LR expression for the efficiency
\begin{align}
\eta  & = \frac{\gamma_1 \gamma_2  \left( \epsilon_1 - \epsilon_2 \right) + 
  \gamma_1 \epsilon_1 }{\gamma_1 \gamma_2  \left( \epsilon_1 -  \epsilon_2\right) - \gamma_2     \epsilon_2 } \;\cdot \; \frac{\epsilon_1 }{\epsilon_2}.
\end{align}
The function $\eta^\star_{\mathrm{LR}}$ can be obtained  using the procedure described in the main text.


\begin{thebibliography}{22}
\bibitem{humphrey} T. E. Humphrey, R. Newbury, R. P. Taylor, and H. Linke, Phys. Rev. Lett. {\bf 89}, 116801 (2002).
\bibitem{benenti}  G. Benenti, K. Saito, and G. Casati, Phys. Rev. Lett. {\bf 106}, 230602 (2011).
\bibitem{yamamoto} K. Yamamoto, O. Entin-Wohlman, A. Aharony, and N. Hatano, Phys. Rev. B {\bf 94}, 121402 (2016).
\bibitem{leggio} B. Leggio and M. Antezza, Phys. Rev. E {\bf 93}, 022122 (2016)
\bibitem{ponmurugan} M. Ponmurugan, arxiv:1604.01912.
\bibitem{indekeu} J. Koning and J. O. Indekeu, Eur. Phys. J. B {\bf 89}, 248 (2016).
\tc{\bibitem{alonso} J. O. Gonz\'alez, D. Alonso, and J. P. Palao, Entropy {\bf 18}, 166 (2016).}
\bibitem{shiraishi} N. Shiraishi, K. Saito, and H. Tasaki, Phys. Rev. Lett. {\bf 117}, 190601 (2016).
\bibitem{ryabov} A. Ryabov and V. Holubec, Phys. Rev. E {\bf 93} 050101 (2016). 
\bibitem{holubec}  V. Holubec and A. Ryabov, J. Stat. Mech.: Th. Exp., 073204 (2016).
\bibitem{armen} A. E. Allahverdyan, K. V. Hovhannisyan, A. V. Melkikh, and S. G. Gevorkian,
Phys. Rev. Lett. {\bf 111}, 050601(2013).
\bibitem{whitney} R. S. Whitney, Phys. Rev. Lett. {\bf 112}, 130601 (2014).
\bibitem{brandner15} K. Brandner and U. Seifert, Phys. Rev. E {\bf 91}, 012121 (2015).
\bibitem{proesmans2} K. Proesmans, B. Cleuren, and C. Van den Broeck, Phys. Rev. Lett. {\bf 116}, 220601 (2016).
\bibitem{bauer} M. Bauer, K. Brandner, and U. Seifert,Phys. Rev. E {\bf 93}, 042112 (2016).
\bibitem{jiang} J.-H. Jiang,  Phys. Rev. E {\bf 90}, 042126 (2014).
\bibitem{polettini} M. Polettini, G. Verley and M. Esposito, Phys. Rev. Lett. {\bf 114}, 050601 (2015).
\bibitem{campisi} M. Campisi and R. Fazio, Nat. Comm. {\bf 7}, 11895 (2016).
\bibitem{shiraishi2} N. Shiraishi, Phys. Rev. E {\bf 92}, 050101 (2015).
\bibitem{bauer2} M. Bauer, D. Abreu, and U. Seifert, J. Physics A: Math. Theor. {\bf 45}, 162001 (2012).
\bibitem{seifert} U. Seifert, Phys. Rev. Lett. {\bf 106}, 020601 (2011).
\bibitem{raz} O. Raz, Y. Suba\c s\i, and R. Pugatch, Phys. Rev. Lett. {\bf 116}, 160601 (2016).
\bibitem{park}  J. S. Lee, H. Park arXiv:1611.07665.
\bibitem{johnson} C. V. Johnson, arXiv:1703.06119.
\tc{
\bibitem{brandner} K. Brandner, K. Saito, and U. Seifert, Phys. Rev. X {\bf 5}, 031019 (2015).
\bibitem{karel} K. Proesmans et al., Phys Rev Lett {\bf 115}, 090601 (2015).
}
\bibitem{entin} O. Entin-Wohlman, J-H. Jiang, and Y. Imry, Phys. Rev. E {\bf 89}, 012123 (2014).
\bibitem{colloqium} G. Benenti, G. Casati, K. Saito, and R. S. Whitney, arXiv:1608.05595.
\bibitem{altaner} B. Altaner, M. Polettini, and M. Esposito, Phys. Rev. Lett. {\bf 117}, 180601 (2016).
\bibitem{brandner13} K. Brandner and U. Seifert, New J. Phys. {\bf 15}, 105003 (2013).
\bibitem{oster} G. Oster, A. Perelson, and A. Katchalsky, Nature {\bf 234}, 393 (1971).
\bibitem{polettiniprojectors} M. Polettini, Lett. Math. Phys. {\bf 105}, 89-107 (2015).
\bibitem{schnak} J. Schnakenberg, Rev. Mod. Phys. \textbf{48}, 571 (1976). 
\bibitem{zia} R. K. P. Zia and B. Schmittmann, J. Stat. Mech, P07012 (2007).
\bibitem{baiesi} M. Baiesi, C. Maes, and B. Wynants, Phys. Rev. Lett. {\bf 103}, 010602 (2009).
\bibitem{verley1} G. Verley, M. Esposito, T. Willaert, and C. Van den Broeck, Nature Comm. {\bf 5}, 4721 (2014).
\bibitem{proesmans1} K. Proesmans, B. Cleuren and C. Van den Broeck, Eur. Phys. Lett. {\bf 109},  20004 (2015).
\bibitem{verley2} G. Verley, T. Willaert, T., C. Van den Broeck, and M. Esposito, Phys. Rev. E {\bf 90}, 052145 (2014).
\bibitem{mahan} G. D. Mahan and J. O. Sofo, Proc. Nat. Acad. Sci. {\bf 93}, 7436 (1996).
\bibitem{svilans} A. Svilans, M. Leijnse, and H. Linke, Comptes Rendus Physique {\bf 17}, 1096 (2016).

\bibitem{esposito}  M. Esposito, R. Kawai, K. Lindenberg and C. Van den Broeck, Phys. Rev. E {\bf 81}, 041106 (2010).
\bibitem{hill} T. L. Hill, J. Theor. Biol. {\bf 10}, 442 (1966).
\tc{ \bibitem{pietzonka}  P. Pietzonka and U. Seifert, arXiv:1705.05817}
{\it Universal trade-off between power, efficiency and constancy in steady-state heat engines}
\bibitem{polettiniphd} M. Polettini, {\it Geometric and Combinatorial Aspects of NonEquilibrium Statistical Mechanics} (Ph.D. dissertation thesis, Alma Mater Studiorum  Universit\`a di Bologna, 2012).
\bibitem{polettini2016} M.  Polettini, G. Bulnes Cuetara, M. Esposito, Phys. Rev. E {\bf 94}, 052117 (2016).
\bibitem{polettiniwip} M. Polettini, work in preparation.
\end{thebibliography}
\end{document}